# 10 Astrophysical Black Holes in the Physical Universe

*Shuang-Nan Zhang*

## CONTENTS

## INTRODUCTION

In modern astronomy, the mystery of black holes (BHs) attracts extraordinary interest for both researchers and the general public. Through the 1930s, the applications of general relativity and quantum mechanics to the studies of the late evolution of stars predicted that stars with different initial masses, after exhausting their thermal nuclear energy sources, may eventually collapse to become exotic compact objects, such as white dwarfs, neutron stars, and BHs. A low-mass star, such as our Sun, will end up as a white dwarf, in which the degeneracy pressure of the electron gas balances the gravity of the object. For a more massive star, the formed compact object can be more massive than around 1.4 solar masses ($M_\odot$), the so-called Chandrasekhar limit, in which the degeneracy pressure of the electron gas cannot resist the gravity, as pointed out by Chandrasekhar. In this case, the compact object has to further contract to become a neutron star, in which most of the free electrons are pushed into protons to form neutrons and the degeneracy pressure of neutrons balances the gravity of the object, as suggested by Zwicky and Landau. Then as Oppenheimer and others noted, if the neutron star is too massive, for example, more than around 3 $M_\odot$, the internal pressure in the object also cannot resist the gravity and the object must undergo catastrophic collapse and form a BH.

Up to now, about 20 BHs with masses around 10 $M_\odot$, called stellar-mass BHs, have been identified observationally. On the other hand, the concept of a BH has been extended to galactic scales. Since the discovery of quasars in the 1960s, these BHs with masses between $10^5$ and $10^{10}$ $M_\odot$, which are called supermassive BHs, are believed to be located in the centers of almost all galaxies. Therefore, tremendous observational evidence supporting the existence of BHs in the Universe is gradually permitting the uncovering of the mysteries of BHs. BH astrophysics has become a fruitful, active, and also challenging frontier research field in modern astrophysics.

Despite tremendous progress in BH research, many fundamental characteristics of astrophysical BHs in the physical Universe remain not fully understood or clarified. In this chapter, I will try to address the following questions: (1) What is a BH? (2) Can astrophysical BHs be formed in the physical Universe? (3) How can we prove that what we call astrophysical BHs are really BHs? (4) Do we have sufficient evidence to claim the existence of astrophysical BHs in the physical Universe? (5) Will all matter in the Universe eventually fall into BHs?

Disclaimer: I will not discuss quantum or primordial BHs. Reviews on theoretical models and observations are intended to be very brief, and thus I will miss many references. Some of the discussions, especially on the question: Will all matter in the Universe eventually fall into BHs?, are quite speculative.

## WHAT IS A BLACK HOLE?

I classify BHs into three categories: *mathematical BHs*, *physical BHs*, and *astrophysical BHs*.

A *mathematical BH* is the vacuum solution of Einstein's field equations of a point-like object, whose mass is completely concentrated at the center of the object, i.e., the singularity point. It has been proven that such an object may possess only mass, angular momentum (spin), and charge, the so-called three hairs. Because of the relatively large strength of the electromagnetic force, BHs formed from gravitational collapse are expected to remain nearly neutral. I therefore discuss only electrically neutral BHs in this chapter. Figure 10.1 is an illustration of the structure of a mathematical BH. The event horizon surrounding the object ensures that no communications can be carried out across the event horizon; therefore, a person outside the event horizon cannot observe the singularity point.

Birkhoff's theorem further ensures that the person outside the event horizon cannot distinguish whether the mass and charge of the object are concentrated at the singularity point or distributed within the event horizon. Therefore, I define a *physical BH* as an object whose mass and charge are all within its event horizon, regardless of the distribution of matter within.

Consequently, a physical BH is not necessarily a mathematical BH. This means that a physical BH may not have a singularity at its center. I further define an *astrophysical BH* as a physical BH that can be formed through astrophysical processes in the physical Universe and within a time much shorter than or at most equal to the age of the Universe. Figure 10.2 is an illustration of a possible process of forming an astrophysical BH through gravitational collapse of matter. So far, all observational studies of BHs have been made on astrophysical BHs. Therefore, the rest of this chapter is focused on them.

## CAN ASTROPHYSICAL BLACK HOLES BE FORMED IN THE PHYSICAL UNIVERSE?

About 70 years ago, Oppenheimer and Snyder studied this problem in their seminal paper "On Continued Gravitational Contraction" (Oppenheimer and Snyder, 1939). Because of the historical and astrophysical importance of this paper, I include a facsimile of the abstract of this paper as Figure 10.3. In the beginning of the abstract, Oppenheimer and Snyder wrote, "When all thermonuclear sources of energy are exhausted a sufficiently heavy star will collapse. Unless . . . [see abstract] this contraction will continue indefinitely." This statement assures that the contraction

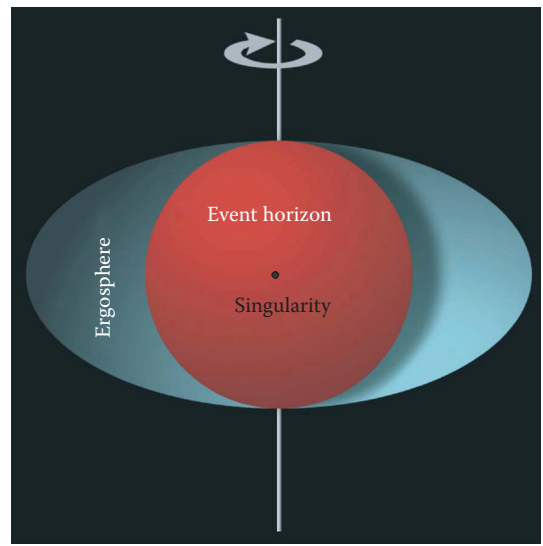

**FIGURE 10.1** Illustration of the structure of a mathematical black hole (BH), which is rotating and has its mass concentrated at its singularity point. The existence of an ergosphere is due to the spin of the BH; a test particle in the ergosphere, although still outside the event horizon, cannot remain stationary. This figure is adapted from artwork in the Wikimedia Commons (available at: http://en.wikipedia.org/wiki/File:Ergosphere.svg).

process illustrated in Figure 10.2 can indeed take place in the physical Universe. In the end of the abstract, Oppenheimer and Snyder arrived at two conclusions that have deeply influenced our understanding of astrophysical BH formation ever since. (1) "The total time of collapse for an observer comoving [called comoving observer in the rest of this chapter] with the stellar matter is finite." This process is depicted in the last frame of Figure 10.2. This is the origin of the widespread and common belief that astrophysical BHs can be formed through gravitational collapse of matter. However, it should be realized that the observer is also within the event horizon with the collapsing matter, once a BH is formed. (2): "An external observer sees the star asymptotically shrinking to its gravitational radius." This means that the external observer will never witness the formation of an astrophysical BH. Given the finite age of the Universe and the fact

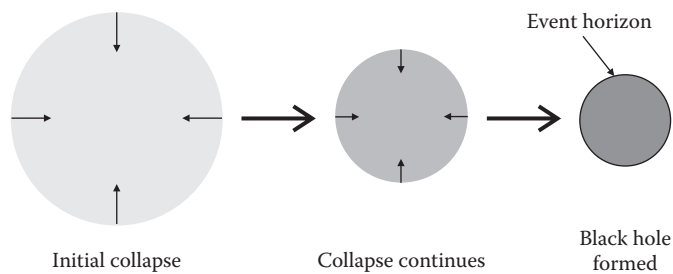

**FIGURE 10.2** Illustration of a possible formation process of an astrophysical black hole (BH). A spherically symmetric cloud of gas collapses under its self-gravity, assuming no internal pressure of any kind. The gas gradually contracts, the size getting smaller and smaller and density getting higher and higher, and eventually falls within the event horizon; it is at this point that a BH is formed. Apparently, not all mass has necessarily arrived at its center at the moment when all matter has just crossed the event horizon; therefore, at least at this moment, this astrophysical BH is just a physical BH and not a mathematical one.

> SEPTEMBER 1, 1939     PHYSICAL REVIEW     VOLUME 56
>
> ## On Continued Gravitational Contraction
>
> J. R. OPPENHEIMER AND H. SNYDER
> *University of California, Berkeley, California*
> (Received July 10, 1939)
>
> When all thermonuclear sources of energy are exhausted a sufficiently heavy star will collapse. Unless fission due to rotation, the radiation of mass, or the blowing off of mass by radiation, reduce the star's mass to the order of that of the sun, this contraction will continue indefinitely. In the present paper we study the solutions of the gravitational field equations which describe this process. In I, general and qualitative arguments are given on the behavior of the metrical tensor as the contraction progresses: the radius of the star approaches asymptotically its gravitational radius; light from the surface of the star is progressively reddened, and can escape over a progressively narrower range of angles. In II, an analytic solution of the field equations confirming these general arguments is obtained for the case that the pressure within the star can be neglected. The total time of collapse for an observer comoving with the stellar matter is finite, and for this idealized case and typical stellar masses, of the order of a day; an external observer sees the star asymptotically shrinking to its gravitational radius.

**FIGURE 10.3** Abstract of the seminal work on astrophysical black hole (BH) formation by Oppenheimer and Snyder (1939). Reprinted with permission from Oppenheimer, J.R. and Snyder, H., *Physical Review*, 56(5), 455–9, 1939. Copyright 1939 by the American Physical Society.

that all observers are necessarily external, the second, and last, conclusion of Oppenheimer and Snyder (1939) seems to indicate that astrophysical BHs cannot be formed in the physical Universe through gravitational collapse.

If, according to Oppenheimer and Snyder, an external observer sees matter asymptotically approach, but never quite cross, the event horizon, then matter must be continually accumulated just outside the event horizon and appear frozen there. Therefore, a gravitationally collapsing object has also been called a "frozen star" (Ruffini and Wheeler, 1971). In fact, the "frozen star" is a well-known novel phenomenon predicted by general relativity, i.e., a distant observer (O) sees a test particle falling toward a BH moving slower and slower, becoming darker and darker, and it is eventually frozen near the event horizon of the BH. This situation is shown in Figure 10.4, in which the velocity of a test particle, as observed from an external observer, approaches zero as it falls toward the event horizon of a BH. This process was also vividly described and presented in many popular science writings (Ruffini and Wheeler, 1971; Luminet, 1992; Thorne, 1994; Begelman and Rees, 1998) and textbooks (Misner et al., 1973; Weinberg, 1977; Shapiro and Teukolsky, 1983; Schutz, 1990; Townsend, 1997; Raine and Thomas, 2005). A fundamental question can be asked: Does a gravitational collapse form a frozen star or a physical BH?

In a recent paper, my student (Yuan Liu) and I summarized the situation as follows (Liu and Zhang, 2009):

> Two possible answers [to the above question] have been proposed so far. *The first one* is that since [the comoving observer] O' indeed has observed the test particle falling through the event horizon, then in reality (for O') matter indeed has fallen into the BH … However, since [the external observer] O has no way to communicate with O' once O' crosses the event horizon, O has no way to "know" if the test particle has fallen into the BH … *The second answer* is to invoke quantum effects. It has been argued that quantum effects may eventually bring the matter into the BH, as seen by O (Frolov and Novikov, 1998). However, as pointed out recently (Vachaspati et al., 2007), even in that case the BH will still take an infinite time to form and the pre-Hawking radiation* will be generated by the accumulated matter just outside the event horizon. Thus this does not answer the question in the real world. Apparently O cannot be satisfied with either answer. In desperation, O may take the attitude of "who cares?" When the test particle is sufficiently close to the event horizon, the redshift is so large that practically no signals from the test particle can be seen by O and apparently the test particle has no way of turning back, therefore the "frozen star" does appear "black" and is an infinitely deep "hole." For practical purposes O may still call it a "BH," whose total mass is also increased by the infalling matter. Apparently this is the view taken by most people in the astrophysical community and general public, as demonstrated in many well-known textbooks (Misner et al., 1973; Hawking and Ellis, 1973; Weinberg, 1977; Shapiro and Teukolsky, 1983; Schutz, 1990; Townsend, 1997; Raine and Thomas, 2005) and popular science writings (Ruffini and Wheeler, 1971; Luminet, 1992; Thorne, 1994; Begelman and Rees, 1998). However when two such "frozen stars" merge together, strong electromagnetic radiations will be released, in sharp contrast to the merging of two genuine BHs (i.e. all their masses are within their event horizons); the latter can only produce gravitational wave radiation (Vachaspati, 2007). Thus this also does not answer the question in the real world.

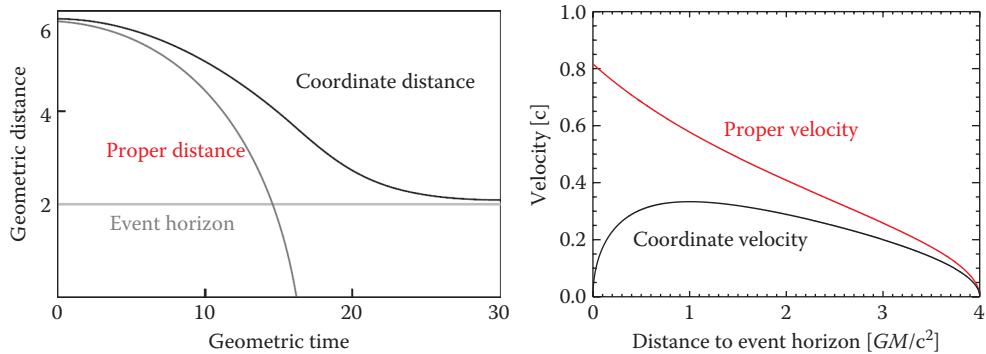

**FIGURE 10.4** Calculation of the motion of a test particle free-falling toward a black hole (BH) starting at rest from $r = 6\ GM/c^2$, where $M$ is the mass of the BH and $c$ is the speed of light in vacuum. Here "proper" and "coordinate" refer to the comoving and external observers, respectively. A set of rigid rulers or milestones are placed everywhere in the system; both the comoving and external observers get the coordinate of the infalling test particle this way. However, the comoving and external observers use their own wristwatches, which are no longer synchronized once the freefall starts. The left panel shows that a test particle takes finite or infinite time to cross the event horizon of the BH, for the comoving and external observers, respectively. The right panel shows that the comoving observer measures the test particle (in fact the observer himself) crossing the event horizon with a high velocity; however, the external observer measures that the test particle stops just outside the event horizon, i.e., is "frozen" to the event horizon. (Left panel adapted from Figure 3 in Ruffini, R. and Wheeler, J.A., *Physics Today*, 30–41,1971. Copyright 1971 by the American Physical Society. With permission.)

The fundamental reason for the above "frozen star" paradox is that the "test particle" calculations have neglected the influence of the mass of the test particle. In reality, the infalling matter has finite mass, which certainly influences the global spacetime of the whole gravitating system, including the infalling matter and the BH. Because the event horizon is a global property of a gravitating system, the infalling matter can cause non-negligible influence to the event horizon. In Figure 10.5, the infalling process of a spherically symmetric and massive shell toward a BH is calculated (Liu and Zhang, 2009) within the framework of Einstein's general relativity. In this calculation, all gravitating mass of the whole system, including both the BH and the massive shell, is taken into account consistently by solving Einstein's field equations. For the comoving observer, the shell can cross the event horizon and arrive at the singularity point within a finite time. *For the external observer, the body of the shell can also cross the event horizon within a finite time but can never arrive at the singularity point*, and its outer surface can only asymptotically approach the event horizon. Compared with the case of the infalling process of a test particle as shown in the left panel of Figure 10.4, the qualitative difference is the expansion of the event horizon as the shell falls in, which does not take place for the test particle case. It is actually the expansion of the event horizon that swallows the infalling shell. Therefore, matter cannot accumulate outside the event horizon of the BH if the influence of the gravitation of the infalling massive shell is also considered.

---

* Hawking radiation is a quantum mechanical effect of black holes (BHs) due to vacuum fluctuations near the event horizon of a BH. The radiation is thermal and blackbody-like, with a temperature inversely proportional to the mass of the BH. Therefore, Hawking radiation is not important at all for the astrophysical BHs we have discussed in this chapter. Pre-Hawking radiation of a BH is in fact not the radiation from the BH, but is hypothesized to come from the matter accumulated just outside the event horizon of the BH. For a remote observer, it may not be possible to distinguish between Hawking radiation and pre-Hawking radiation (even if it does exist) unless we know precisely the properties of the BH and the matter accumulated just outside its event horizon.

The calculations shown in Figure 10.5 still neglected one important fact for real astrophysical collapse. There is always some additional matter between the observer and the infalling shell being observed (we call it the inner shell), and the additional matter is also attracted to fall inward by the inner shell and the BH. We thus modeled the additional matter as a second shell (we call it the outer shell) and calculated the motion of the double-shell system. Our calculations show that in this case the inner shell can cross the event horizon completely even for the external observer, but it can still never arrive at the central singularity point (Liu and Zhang, 2009). Based on these calculations, we can conclude that *real astrophysical collapses can indeed form physical BHs*, i.e., all mass can cross the event horizon within a finite time for an external observer, and thus no "frozen stars" are formed in the physical Universe. A rather surprising result is that matter can never arrive at the singularity point, according to the clock of an external observer. This means that *astrophysical BHs in the physical Universe are not mathematical BHs because, given the finite age of the Universe, matter cannot arrive at the singularity point* (Liu and Zhang, 2009). This justifies my classifications of BHs into three categories.

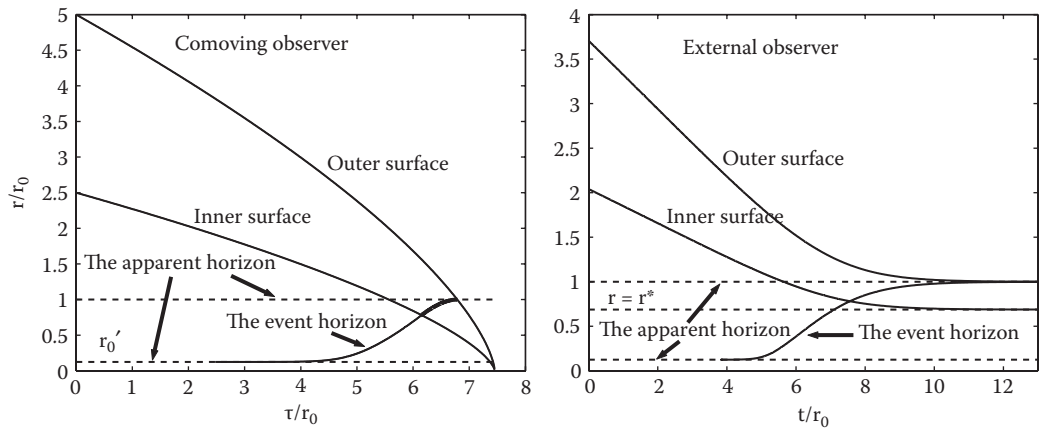

**FIGURE 10.5** The infalling process of a spherically symmetric and massive shell toward a black hole (BH), calculated within the framework of Einstein's theory of general relativity. The left and right panels show the observations made by a comoving observer and an external observer, respectively; the two solid lines mark the inner and outer surfaces of the shell, respectively. The expansion of the event horizon as the shell falls in is also shown. For the comoving observer, the shell can cross the event horizon and arrive at the singularity point within a finite time. For the external observer, the body of the shell can also cross the event horizon within a finite time, but it can never arrive at the singularity point, and its outer surface can only asymptotically approach the event horizon. (This figure is adapted from panels (a) and (b) of Figure 2 in Liu, Y. and Zhang, S.N., *Physics Letters B*, 679, 88–94, 2009. Copyright 2009 by Elsevier. With permission.)

## HOW CAN WE PROVE THAT WHAT WE CALL ASTROPHYSICAL BLACK HOLES ARE REALLY BLACK HOLES?

The defining characteristic of an astrophysical BH is that all its gravitating mass is enclosed by its event horizon, and consequently all infalling matter will fall into its event horizon within a finite time of an external observer. Therefore, it has been commonly believed that the final and unambiguous confirmation of the detection of BHs requires direct evidence for the existence of the event horizon of a BH. However, by virtue of the very definition of the event horizon that no light can escape from it to infinity, direct evidence for the existence of the event horizon of a BH can never be obtained by a distant observer. However, in science direct evidence is not always what leads to the discovery of something. For example, we never "see" directly many particles created in accelerator experiments, whose existence is usually inferred by their decay products. Actually, quarks do not even exist in free forms, and very few scientists today question that quarks exist. Searching for dark matter* particles, which may be created in CERN's Large Hadron Collider (LHC) experiments, is currently under way.

However, even if dark matter particles are being produced there, these particles have no chance of annihilating or even interacting in these detectors. Therefore, only indirect evidence, such as "missing mass," can be used to demonstrate detection of dark matter particles in accelerator experiments. In astronomy, similar situations exist. For example, no "direct" evidence exists for dark matter and dark energy in the Universe. However, dark matter and dark energy are widely believed to exist, from a collection of many pieces of indirect evidence. Do we have a collection of indirect evidence to prove that what we call astrophysical BHs are really BHs? Because in astronomy we are dealing with astrophysical BHs with masses over a range of at least eight orders of magnitude and located in very different astrophysical environments, here I suggest five criteria, or parameters, in determining whether astronomers have found astrophysical BHs:

1. The concept and theoretical model based on astrophysical BHs can be used to explain a series of common observational phenomena known previously.
2. The same concept and theoretical model based on astrophysical BHs can be used to explain the ever-increasing volume of new observational phenomena.
3. No counterevidence comes forward against the model based on astrophysical BHs.
4. The BH formation and evolution scenario inferred from those observational phenomena are self-consistent and physically and astrophysically reasonable.
5. There is no alternative theoretical model that can also explain the same or even more phenomena with the same or even better success than the astrophysical BH model.

Although general, the above five criteria meet the highest standard for recognizing new discoveries in experimental physics and observational astronomy. As a matter of fact, these criteria also meet Carl Sagan's principle that "extraordinary claims require extraordinary evidence" because of the importance and impacts of discovering BHs in the Universe. Indeed, it is debatable that the discoveries of very few, if any, astrophysical objects meet such stringent and extensive requirements.

## DO WE HAVE SUFFICIENT EVIDENCE TO CLAIM THE EXISTENCE OF ASTROPHYSICAL BLACK HOLES IN THE PHYSICAL UNIVERSE?

Having given up the hope of finding "direct" evidence for the existence of the event horizon of a BH, we must search for other supporting evidence for the existence of BHs, following the five criteria I proposed in the previous section. The next hope is to study what happens when matter or light gets sufficiently close to or even falls into BHs and then explains in this way as many observational phenomena as possible. Around a BH, several important effects might be used to provide indirect evidence for the existence of the BH:

1. The surface of a BH or matter hitting it does not produce any radiation detectable by a distant observer; this is a manifestation of the event horizon of a BH.
2. There exists an innermost stable circular orbit for a BH, beyond which matter will free-fall into the BH; this orbital radius is a monotonic function of the angular momentum of a BH, as shown in Figure 10.6. In some cases this general relativistic effect can be used to measure the spin of a BH, for example, by fitting the continuum spectrum or relativistically blurred lines produced from the inner region of an accretion disk around a BH (Loar, 1991; Zhang et al., 1997).
3. The very deep gravitational potential around a BH can produce strong gravitational lensing effects; an isolated BH may be detected this way.
4. The very deep gravitational potential around a BH can cause matter accreted toward a BH to convert some of its rest mass energy into radiation; an accreting BH may be detected this way. In Figure 10.7, I show the conversion efficiency of different kinds of BH accretion systems, in comparison with the conversion efficiencies of other astrophysical systems.
5. For a spinning BH, its ergosphere (as shown in Figure 10.1) will force anything (including magnetic field lines) within the ergosphere to rotate with it; the Penrose or magnetic Penrose mechanism may allow the spin energy of a BH to be extracted to power strong outflows (Blandford and Znajek, 1977). Sometimes outflows can also be produced from accretion disks around non-spinning BHs (Blandford and Payne, 1982).

---

* This is a kind of matter believed to dominate the total mass of the Universe, but it does not produce any electromagnetic radiation. For details on dark matter, please refer to Bloom's chapter in this volume.

## Luminous Accreting Black Holes

If there is a sufficient amount of matter around a BH, matter under the gravitational attraction of the BH will be accreted toward it, and in this process an accretion disk can be formed surrounding the BH. Under certain conditions a geometrically thin and optically thick accretion disk can be formed (Shakura and Sunyaev, 1973), which is very efficient in converting the gravitational potential energy into thermal radiation. The radiative efficiency (ratio between radiated energy and the rest mass energy of accreted matter) is approximately inversely proportional to the inner boundary radius of the accretion disk, as shown in Figure 10.7, because the matter between the inner disk boundary and the event horizon of the BH is free-falling, and almost all the kinetic energy is carried into the BH. Please refer to the caption of Figure 10.7 for detailed explanations.

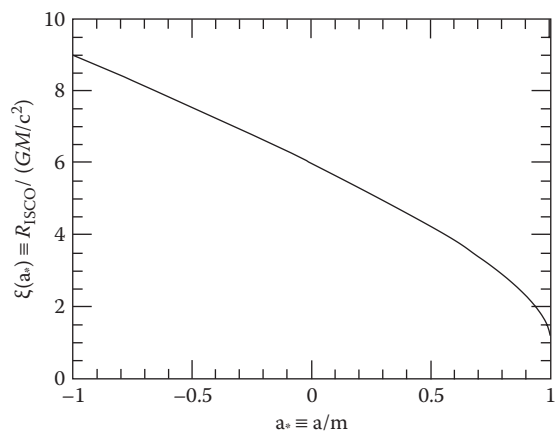

**FIGURE 10.6** The radius of the innermost stable circular orbit ($R_{ISCO}$) of a black hole (BH) as a function of the spin parameter ($a_*$) of the BH, i.e., the dimensionless angular momentum; a negative value of $a_*$ represents the case that the angular momentum of the disk is opposite to that of the BH. The spin angular momentum of a BH, the seond parameter for a BH, can be measured by determining the inner accretion disk radius if the inner boundary of the disk is the innermost stable circular orbit of the BH.

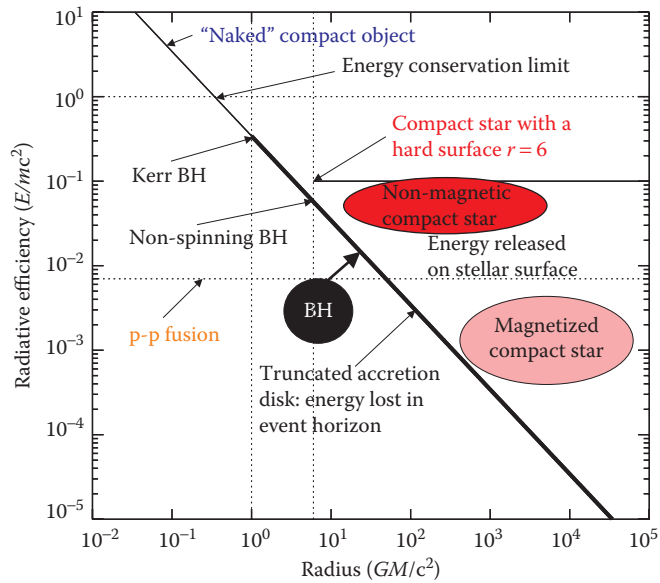

**FIGURE 10.7** For an accretion disk around a black hole (BH), the radiative efficiency (ratio between radiated energy and the rest mass energy of accreted matter) is approximately inversely proportional to the inner boundary radius of the accretion disk. Here it is assumed that the radiation produced by the accreted matter (in almost free fall) between the disk boundary and the event horizon of the BH is negligible; however, the radiative efficiency is slightly higher if the very weak emission from the matter between the disk boundary and the event horizon of the BH is also considered (Mahadevan, 1997; also see the caption for Figure 10.9). The diagonal line shows a $1/r$ scaling, calibrated to take the value of 0.057 when $r = 6$. The thick black line is for strongly suspected BH accreting systems. The range of $r = 1$–$9$ corresponds to the innermost stable circular orbit of a BH with different spin, assuming that the disk extends all the way there; the radiative efficiency ranges from a few to several tens of percent, far exceeding the p-p fusion radiative efficiency taking place in the Sun. The case for $r > 9$ corresponds to a truncated accretion disk, whose radiative efficiency can be extremely low, because energy is lost into the event horizon of the BH. The thin solid black horizontal line is for the 10% efficiency when matter hits the surface of a neutron star where all gravitational energy is released as radiation. The thin solid black diagonal line above the point marked for "Kerr BH" (Kerr black hole) is for a speculated "naked" compact object whose hose surface radius is extremely small, and thus the radiative efficiency can be extremely high.

Figure 10.8 describes accreting disks surrounding a Kerr (spinning) BH (*left*) and a Schwarzschild (non-spinning) BH (*right*); the inner boundary of the disk stops at the innermost stable circular orbit of the BH when the accretion rate is around 10% of the Eddington rate. Such high radiation efficiency is commonly observed in the luminous state of a binary system suspected to contain a BH of several solar masses as the accretor (Remillard and McClintock, 2006), or in a quasi-stellar object (QSO) (also called a quasar or active galactic nucleus [AGN]) suspected to harbor at the center of a galaxy a supermassive BH of millions to billions of solar masses as the accretor (Yu and Tremaine, 2002). The BH accretion model, with essentially only three parameters (two for the mass and spin of a BH, and one for the accretion rate of the disk), can explain the many observed properties of dozens of BH binary systems in the Milky Way and countless AGNs in the Universe (Zhang, 2007b). Currently, no single alternative model can be used in a systematic and consistent way to explain these same observations in those binary systems and AGNs.

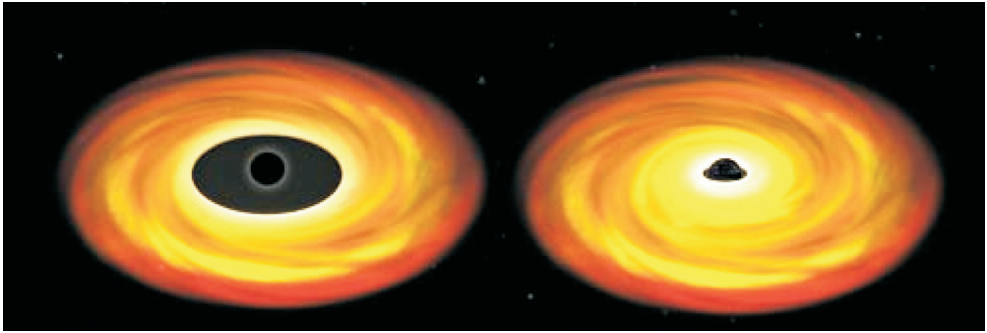

**FIGURE 10.8** Accretion disks around non-spinning (*left*) and spinning (*right*) black holes (BHs). For the spinning BH, both its inner disk and event horizon radii are smaller, thus providing a deeper gravitational potential well for a more efficient energy conversion, reaching a maximum efficiency of about 42% (Page, D.N., and Thorne, K.S., *Astrophysical Journal*, 191, 499–506, 1974). (Courtesy of NASA/CXC/M. Weisskoff http://chandra.harvard.edu/photo/2003/bhspin/.)

### Faint Accreting Black Holes

When the radiation of the disk is substantially below 10% Eddington luminosity, the optically thin and geometrically thick disk tends to retreat away from the BH, and the central region is replaced by some sort of radiatively inefficient accretion flow, for example, the advection-dominated accretion flow (Narayan and Yi, 1994). Generically, this corresponds to the case for $r > 9$ in Figure 10.7, i.e., a truncated accretion disk, whose radiative efficiency can be extremely low because almost all gravitational potential energy is converted into the kinetic energy of the accreted matter that free-falls into the BH and thus is lost into the event horizon of the BH. This model has been used to explain the extremely low luminosity of the quiescent state of BH binaries (Shahbaz et al., 2010), the inferred supermassive BHs in the center of the Milky Way, and many nearby very-low-luminosity AGNs (Ho, 2008); normally, $r > 100$ for these extremely underluminous systems. Recently, evidence has been found for the truncation radius in the range of $r = 10$–100 for binary systems in their normal, but slightly less luminous, states, for example, around 0.01 to 0.1 Eddington luminosity (Gierliński et al., 2008). The top panel of Figure 10.9 shows a theoretical calculation of the expected truncation radius as a function of accretion rate $\dot{M}$ (Liu and Meyer-Hofmeister, 2001), i.e., roughly $r \propto \dot{M}^{-1/2}$. The bottom panel of Figure 10.9 shows the observed accretion disk luminosity $L$ as a function of observationally inferred disk truncation radius (Shahbaz et al., 2010), i.e., roughly $L \propto r^{-3}$. Therefore, the radiative efficiency $\eta = L/\dot{M} \propto r^{-3} r^2 \propto 1/r$, as shown in Figure 10.7. Once again, the BH accretion disk model is so far the only one that can explain all these observations across huge dynamic ranges of mass, time, space, environment, and luminosity.

### The Supermassive Black Hole at the Center of the Milky Way

A single strong case for a BH lies at the center of the Milky Way. As shown in the top panel of Figure 10.10, the mass of the central object is measured to be around 4 million solar masses by observing the stellar motions very close to it; the closest distance between the S2 star (the currently known nearest star to the center) and the center is around 2,100 times the radius of the event horizon of the BH, thus excluding essentially all known types of single astrophysical objects as the compact object there. The bottom panel of Figure 10.10 shows the extremely compact size of the radio signal-emitting region, which is merely several times the radius of the event horizon of the BH, ruling out a fermion star model and also disfavoring a boson star model. In fact, the extremely low radiation efficiency of this object requires that the central object cannot have a surface, i.e., the majority of the gravitational energy is converted to the kinetic energy of the accreted matter and subsequently lost into the BH (Broderick et al., 2009). Putting all these pieces of supporting evidence together does not leave much room for a non-BH object as the central compact object of the Milky Way. The properties of this system can be well explained with the same BH accretion model used to explain the quiescent-state properties of other low-luminosity AGNs and galactic BH binaries (Yuan et al., 2003).

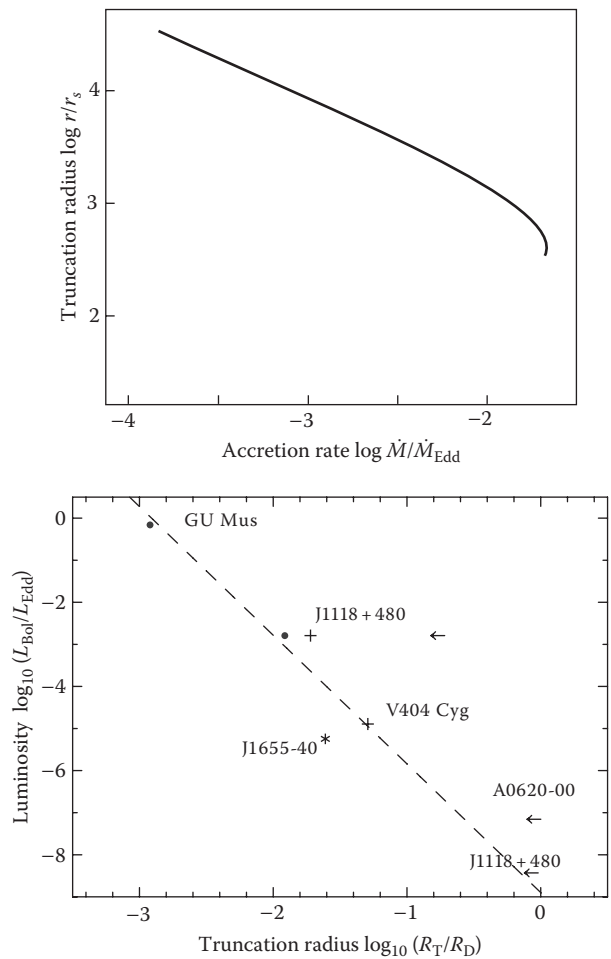

**FIGURE 10.9** Accretion disk truncation radius and luminosity for faint (low-luminosity) accreting black holes (BHs). The top panel presents a theoretical calculation of the expected truncation radius (normalized to the radius of the event horizon of a BH) as a function of accretion rate (normalized to the Eddington rate). The bottom panel presents accretion disk luminosity (normalized to the Eddington luminosity) as a function of the observationally inferred disk truncation radius (normalized to an arbitrary unit). (The top and bottom panels give roughly $r \propto \dot{M}^{-1/2}$ and $L \propto r^{-3}$, respectively.* Therefore, the radiative efficiency is $\eta = L/\dot{M} \propto r^{-3} r^2 \propto 1/r$, as shown in Figure 10.7. The data points on the bottom panel are for suspected BH accretion systems. (top panel adapted from Figure 1 in Liu, B.F. and Meyer-Hofmeister, E., *Astronomy and Astrophysics*, 372, 386–90, 2001. Copyright (2001) by *Astronomy and Astrophysics*. With permission. Bottom panel adapted from Figure 8 in Shahbaz, T., Dhillon, V.S., Marsh, T.R., et al., *Monthly Notices of the Royal Astronomical Society*, 403, 2167–75, 2010. Copyright (2010) by *Wiley*. With permission.) (*More precisely, the top panel gives $r \propto \dot{M}^{-2/3}$, thus $\eta = L/\dot{M} \propto r^{-3} r^{3/2} \propto r^{-3/2}$, consistent with the prediction of the advection-dominated accretion flow model if the emission between the disk boundary and the event horizon of the BH is not negligible [Mahadevan, 1997].)

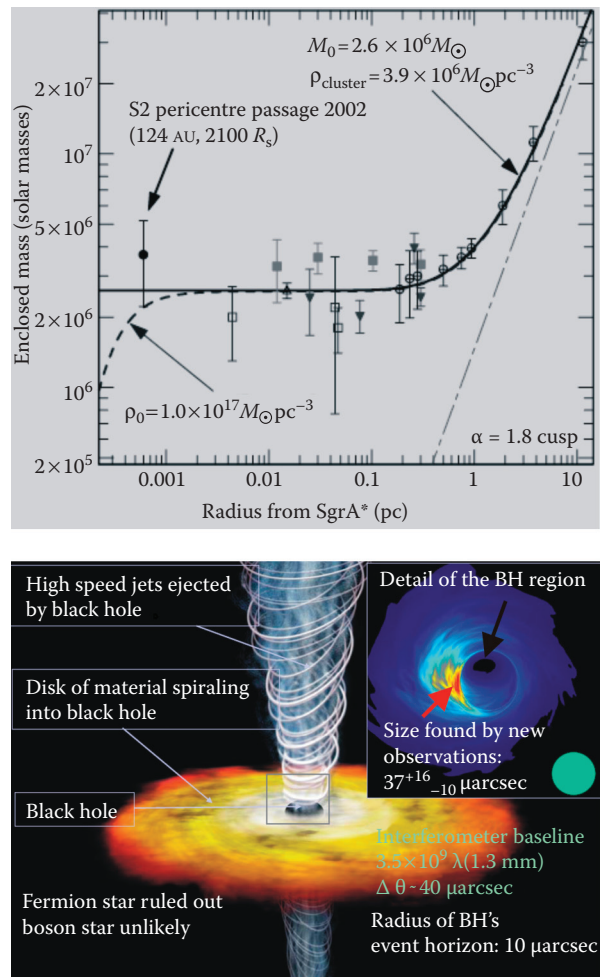

**FIGURE 10.10**   Mass and size of the supermassive black hole (BH) at the center of the Milky Way. The top panel shows the enclosed mass as a function of radius from the dynamic center of the Milky Way. The bottom panel illustrates our current understanding of what is going on around the suspected supermassive BH at the center of the Milky Way. The inset at the upper-right corner of the bottom panel shows that the angular resolution of the observation is about 40 µarcsec (marked as the green circular area), obtained with the $\lambda = 1.3$ mm wavelength interferometer with a baseline of $3.5 \times 10^9$. The inferred size of the radio signal-emitting region (red arrow) is about 37 µarcsec, comparable to the size of the event horizon of this supermassive BH, which is about 10 µarcsec (black arrow). This suggests that the compact object must be at least smaller than several times the size of the event horizon of the suspected supermassive BH, thus ruling out a fermion star model and disfavoring a boson star model. (Top panel reprinted from Schödel, R., Ott, T., Genzel, R., et al., *Nature*, 419, 694–6, 2002. Copyright (2002) by Macmillan Publishers Ltd. With permission. Bottom panel adapted from the online supplementary material of Doeleman, S.S., Weintroub, J., Rogers, A.E.E., et al., *Nature*, 455, 78–80, 2008. Copyright (2008) by Macmillan Publishers Ltd. With permission.)

## Comparison with Accreting Neutron Stars

The thin solid black horizontal line in Figure 10.7 is for the 10% efficiency when matter hits the surface of a neutron star where all gravitational energy is released as radiation. Essentially, all accreted matter can reach the surface of a neutron star if the surface magnetic field of the neutron star is so low that the magnetic field pressure does not play a significant role in blocking the accreted matter from reaching the surface of the neutron star; in this case, the radiation efficiency is not much below 10%. However, the radiation efficiency can be substantially below 10%, when the accretion rate is very low such that the surface magnetic field of the neutron star can block the accreted matter through the so-called propeller effect (Zhang et al., 1998). If the accretion disk around the neutron star at very low accretion rate is in the advection-dominated flow state, some of the accreted matter can still reach the surface of the neutron star and produce a non-negligible amount of radiation from the surface of the neutron star (Zhang et al., 1998; Menou et al., 1999). Therefore, for two binary systems with a BH and a neutron star as the accretors, respectively, of material from a normal star, the neutron star binary will appear brighter, even if their accretion disks are exactly the same, as shown in Figure 10.7. This expectation has been observationally confirmed for all known BH and neutron star binaries at their quiescent states as shown in Figure 10.11 (Narayan and McClintock, 2008). Therefore, the simple accreting BH (and neutron star) model can explain nicely a large collection of observations.

## Isolated Black Holes

Clearly, for an isolated astrophysical BH, which is not surrounded by dense medium and thus is not actively accreting matter, the only way to detect it is through a gravitational lensing effect (Paczynski, 1986, 1996). So far, several candidate BHs have been found this way (Bennett et al., 2002; Mao et al., 2002). However, practically speaking, lensing observations can find only candidate BHs because it is extremely difficult to exclude all other possibilities responsible for the detected lensing events. Additional evidence supporting the BH nature of the candidate object must be sought, e.g., x-ray emission from accreted interstellar medium onto the putative BH (Agol and Kamionkowski, 2002). Currently, only an upper limit on the anticipated x-ray emission from one candidate has been observed, indicating that the radiative efficiency is as low as around $10^{-10}$—$10^{-9}$, assuming that the putative BH is located in the normal interstellar medium (ISM) (Nucita et al., 2006); this efficiency is far below the range shown in Figure 10.7. However, as recently found, all microquasars are located in parsec-scale cavities with density lower by at least three orders of magnitude than the normal ISM (Hao and Zhang, 2009). Then the estimated radiative efficiency upper limit might be increased by at least three orders of magnitude, if this putative BH is also located in a very-low-density cavity. Even in this case, the radiative efficiency would still be in the lowest end in Figure 10.7, thus indicating that the majority of the kinetic energy of the accreted matter is lost into the event horizon of the BH.

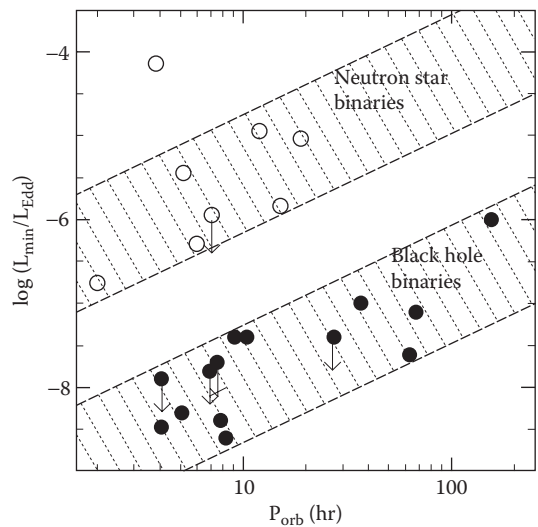

**FIGURE 10.11** Comparison of the quiescent bolometric luminosity between neutron star and black hole (BH) binaries. In an accreting binary, its quiescent-state luminosity (lowest luminosity state) is scaled positively with its compact object mass and orbital period, regardless of whether the accretor is a neutron star or a BH. The main difference between neutron star and BH accretors is that the surface radiation of the neutron star makes the neutron star system brighter (in units of the Eddington luminosity) for the same orbital period. (Adapted from Narayan, R. and McClintock, J.E., *New Astronomy Reviews*, 51, 733–51, 2008. Copyright 2008 by Elsevier. With permission.)

## Luminous "Naked" Compact Objects?

In Figure 10.7, the thin solid black diagonal line above the point marked for "Kerr BH" is for a speculated "naked" compact object, whose surface radius is extremely small but not enclosed by an event horizon. The concept for a "naked" compact object is related to "naked" singularity, which is not enclosed by an event horizon; a "naked" singularity can be formed in a variety of gravitational collapse scenarios (Pankaj, 2009), thus breaking Penrose's cosmic censorship.* A key characteristic for an accreting "naked" singularity is that radiation can escape from it, in sharp contrast to an accreting BH, as illustrated in Figure 10.12. Following the arguments I made when answering the question, Can astrophysical BHs be formed in the physical Universe?, "naked" compact objects, rather than "naked" singularities, might be formed in the physical Universe. In this case, the radiative efficiency can be very high, depending on the radius of the "naked" compact object. For extremely small radii, the efficiency may exceed 100%, implying that the energy of the "naked" compact object is extracted. Unfortunately, so far there has been no observational evidence supporting this conjecture. However, this possibility, if true, may have fundamental impacts regarding the evolution and fate of the Universe, as I will discuss at the end of this chapter.

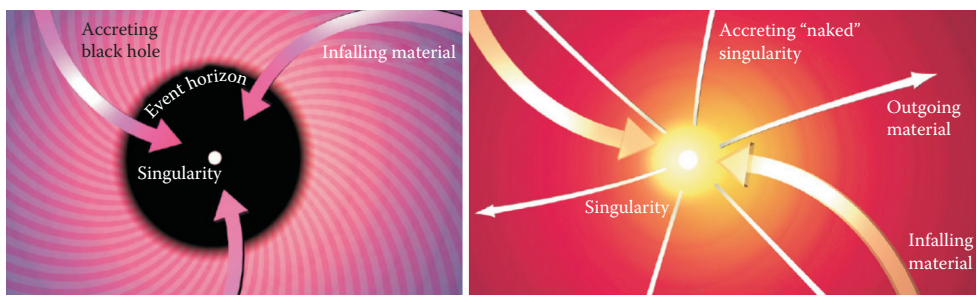

**FIGURE 10.12** Comparison between an accreting black hole (*left*) and an accreting "naked singularity" (*right*), which can be luminous for a distant observer. (Adapted from the online slides of *Scientific American* (available at http://www.scientificamerican.com/slideshow.cfm?id=naked-singularities&photo_id=DC1F7444-DCC7-F2E4-2EF03074D470B687 and http://www.scientificamerican.com/slideshow.cfm?id=naked-singularities&photo_id=DC1F8C9A-0E60-3C59-5CD90FA1B4505784). Copyright (2009) by Alfred T. Kamajian. With permission.)

---

* Penrose's cosmic censorship conjectures that each and every singularity in the Universe is enclosed by an event horizon, i.e., there is no "naked" singularity in the Universe.

## Relativistic Jets

A spinning BH can also power relativistic jets, as observed commonly from AGNs (or quasars) and galactic BH binaries (or microquasars), as shown in Figure 10.13. This can happen when large-scale magnetic fields are dragged and wound up by the ergosphere (see Figure 10.1) of a spinning BH, as shown in Figure 10.14. The twisted and rotating magnetic field lines can then accelerate the infalling plasmas outward along the spin axis of the BH to relativistic speeds (Blandford and Znajek, 1977), producing powerful relativistic jets that can carry a substantial amount of the accretion power and travel to distances far beyond these binary systems or their host galaxies. Recent studies have shown that the BHs in microquasars are indeed spinning rapidly (Zhang et al., 1997; Mirabel, 2010; McClintock et al., 2009). Once again, a conceptually simple BH accretion model can explain the observed relativistic jets from accreting BH systems with very different scales.

## Gamma-Ray Bursts

Gamma-ray bursts (GRBs) (Klebesadel et al., 1973; Fishman and Meegan, 1995; Gehrels et al., 2009) are strong gamma-ray flashes with an isotropic energy between $10^{50}$ and $10^{54}$ ergs released in seconds or shorter for each event. They originate at redshifts as high as 8.3 (Salvaterra et al., 2009; Tanvir et al., 2009) or even beyond 10 (thus seen as they were at only a few percent of the age of the Universe [Lin et al., 2004]). GRBs are the biggest explosions in the Universe since the Big Bang and can be used to probe the evolution of the Universe. At least some of the "long" GRBs, with duration approximately more than 2 sec, are believed to be produced from spinning BHs accreting at extremely high rates (Gehrels et al., 2009; Mézáros, 2009; Zhang, 2007a). In this picture, a spinning BH is formed as a massive star ends its life in a gravitational collapse; the fallback matter after the accompanying supernova (SN) explosion forms an accretion disk around the BH. In an extremely violent process similar to that shown in Figure 10.14, super-relativistic super-relativistic jets, with Lorenz factors of hundreds to thousands, are produced, which produce luminous and also highly beamed gamma-ray emissions.*

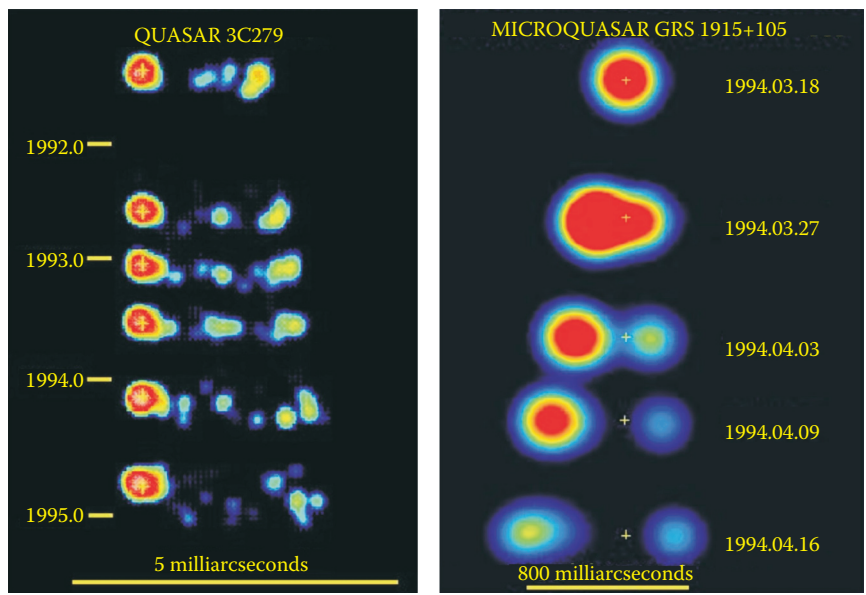

**FIGURE 10.13**  Relativistic jets from the quasar 3C 279 (*left panel*: an active galactic nucleus with a redshift of $z = 0.536$) and the microquasar GRS 1915+105 (*right panel*: a Galactic black hole [BH] binary). The radio images from top to bottom are observed sequentially at different times; in the left panel, the starting time of each year is marked as a short bar, and in the right panel the date when each observation was made is shown. Radio signals are synchrotron radiation from entrained particles in higher-density portions of the jets illustrated elsewhere in this chapter. The crosses mark the locations of the BHs, providing reference points for measuring the proper motions of jets. The lengths of the long horizontal bars (5 and 800 mas in the left and right panels, respectively) near the bottom of each panel show the angular size scales of the jets on them. The Galactic object (right panel) shows a two-sided jet; the color scale uses redder colors for higher intensity. The jet coming toward us is relativistically Doppler boosted and thus is brighter than the counter jet. The quasar is at cosmological distance; counter jets are not normally observed in such cases because they are very faint. The inferred intrinsic velocities of the jets for both systems are more than 98% of the speed of the light. (Adapted from Mirabel, I.F. and Rodriguez, L.F., *Nature*, 371, 46–8, 1994. Copyright 1994 by Macmillan Publishers Ltd. With permission.)

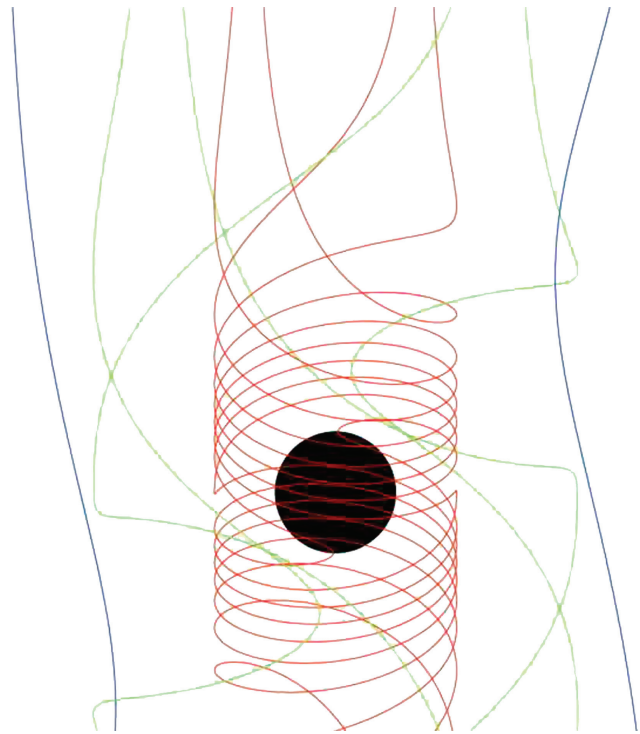

**FIGURE 10.14** Illustration of the production process of a relativistic jet, similar to that shown in Figure 10.10, by an accreting spinning black hole (BH). The magnetic field lines are wound up by the ergosphere (see Figure 10.1) of the spinning BH, because nothing can stay stationary there and must rotate with the spinning BH. Accreted matter into this region is spun out with relativistic speeds along the spin axis of the BH, because the accreted matter is fully ionized and must move along these wound up magnetic field lines. (Reprinted from Figure 4d in Meier et al. [2001]. With permission from the American Association for the Advancement of Science.)

### PUTTING IT ALL TOGETHER: ASTROPHYSICAL BLACK HOLES HAVE BEEN DETECTED

Therefore, the BH accretion (and outflow) model can be used to explain a vast array of astrophysical phenomena across huge dynamical ranges of time, space, mass, luminosity, and astrophysical environments.

The first collection of "indirect" evidence for the existence of BHs is with the radiative efficiency when matter falls toward a central compact object. As we have proven (Liu and Zhang, 2009), matter in a gravitational potential well must continue to fall inward (but cannot be "frozen" somewhere), either through the event horizon of a BH or hitting the surface of a compact object not enclosed by an event horizon but with a radius either larger or smaller than the event horizon of the given mass (called a compact star or "naked" compact object, respectively). No further radiation is produced after the matter falls through the event horizon of the BH; thus, the majority of the kinetic energy of the infalling matter is carried into the BH. On the other hand, surface emission will be produced when matter hits the surface of the compact star or "naked" compact object, because it is not a BH. Therefore, the radiative efficiencies for these different scenarios are significantly different, as shown in Figure 10.7. Currently, all observations of the strongly suspected accreting BHs in binary systems or at the centers of many galaxies agree with the BH accretion model, over a huge range of accretion rates.

---

* For more details on supernovae and gamma-ray bursts, please refer to the chapter by Filippenko in this volume.

The second collection of "indirect" evidence for the existence of BHs is with the relativistic jets from microquasars (accreting BH binaries), quasars (accreting supermassive BHs), and GRBs (also called collapsars, i.e., accreting BHs just formed in a special kind of SN event). In Figure 10.15, a unified picture of BH accretion and outflow is presented for these three seemingly very different kinds of systems. The key ingredient of the model is that the combination of the deep gravitational potential well and the ergosphere of a spinning BH extracts both the potential energy and the spinning energy of the BH, producing strong electromagnetic radiation and powerful relativistic outflows. This model explains current observations satisfactorily.

Among all competing models (many of them can only be used to explain some of these phenomena), the BH accretion (and outflow) model is the simplest, and the astrophysical BHs are also the simplest objects, with only two physical properties (mass and spin). The BH masses and spin parameters, found by applying the BH accretion model to many different kinds of data, are physically and astrophysically reasonable and also well understood so far. The mass of a stellar-mass BH comes from the gravitational collapse of the core of a massive star and the subsequent matter in-falling process; some of the core-collapse supernovae and GRBs are manifestations of this process. A supermassive BH grows up by accreting matter in its host galaxy; the active accretion process makes the galaxy show up as a QSO. The BH accretion process can efficiently increase the spin of a BH, by transferring the angular momentum of the accreted matter to the BH.

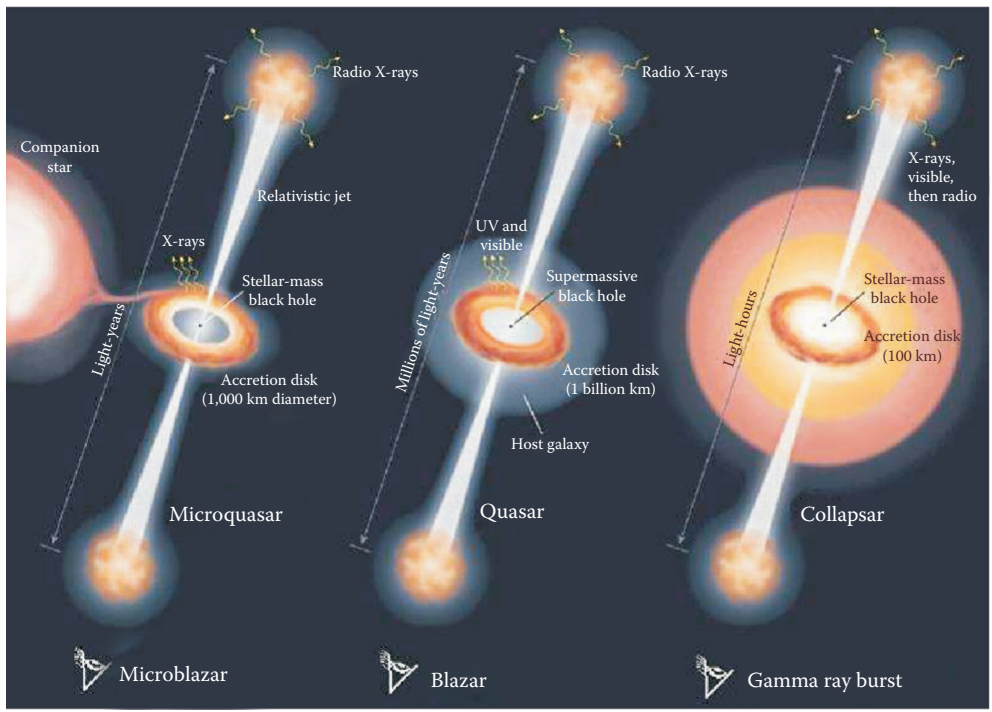

**FIGURE 10.15** Unified picture of black hole accretion and outflow model for three kinds of astrophysical systems with very different observational characteristics and extremely different scales of mass, time, size, luminosity, and astrophysical environments. (Reprinted from Mirabel and Rodriguez, *Sky & Telescope*, p. 32, 2002. With permission.)

Is the model falsifiable? If surface emission is detected from the putative BH in any of the above systems, one can then confidently reject the validity of the BH accretion model, at least for that specific system. For the only other two kinds of compact objects known, i.e., white dwarfs and neutron stars, surface emissions have been commonly detected. Yet so far this has not happened to any of the putative accreting BH systems we discussed above. Therefore, there is no counterevidence against the BH accretion model used to explain all phenomena discussed in this chapter.

Positive identification of astrophysical BHs in those objects also satisfies the principle of Occam's razor, i.e., that "entities should not be multiplied unnecessarily," commonly interpreted as "take the simplest theory or model among all competitors." However, the history of science tells us that Occam's razor should be used only as a heuristic to guide scientists in the development of theoretical models rather than as an arbiter between published models; we eventually accept only models that are developed based on existing data but can also make falsifiable predictions, are confirmed with additional data, and can explain new data or new phenomena. This is indeed what has happened to the BH accretion model. In this sense, the BH accretion (and outflow) model has survived all possible scrutiny.

I therefore conclude that *we now have sufficient evidence to claim that we have found astrophysical BHs, at least in some galactic binary systems, at the center of almost every galaxy, and as the central engines of at least some long GRBs.*

## WILL ALL MATTER IN THE UNIVERSE EVENTUALLY FALL INTO BLACK HOLES?

In the previous sections, I have emphasized the importance of BH accretion and actually relied on the BH accretion model to argue in favor of the existence of astrophysical BHs in the physical Universe. It is then not accidental to ask the following question: Will all matter in the Universe eventually fall into BHs? As a matter of fact, I have indeed been asked this question numerous times by nonprofessional researchers when I gave public talks on BHs; somehow only the professional researchers hesitate to ask this question. Each time I have almost randomly used one of three answers: "yes," "no," or "I don't know." Here I attempt to provide some rather speculative discussions on this question.

Ignoring the Hawking radiation of a BH and assuming that no "naked" singularities (compact stars) exist in the physical Universe (i.e., that Penrose's cosmic censorship holds), indeed it is inevitable that all matter (including dark matter and perhaps all forms of energy) will eventually fall into BHs if the Universe is not expanding (i.e., is stationary) and does not have a boundary. This is because regardless of how small the probability is for a particle or a photon to fall into a BH, it eventually has to fall into a BH after a sufficiently large number of trials. A universe made of only BHs is of course an eternally dead universe. An eternally expanding universe will save some matter from falling into BHs because eventually particles or even light escaping from a galaxy or those (such as dark matter and hot baryons and electrons) that are already in intergalactic regions may never reach another galaxy and thus not fall into any BH. However, whatever is left in a given galaxy will still eventually fall into one of the BHs in the galaxy. Therefore, each galaxy will be made of only BHs, and these BHs may collide with one another to become a huge BH. It is inevitable that in the end each galaxy will be just a huge

BH. Then eventually the expanding universe will be made of numerous huge BHs moving apart from one another, with some photons and particles floating between them and never quite catching them. This universe is still a dead one. If at some point the universe begins to contract, then particles (including dark matter) and photons outside BHs will begin to be sucked into BHs, and BHs will also begin to merge with each other. Eventually, the whole universe may become just a single huge BH.

Can the Hawking radiation intervene to rescue our Universe from an eternal death? It is easy to calculate that for a 10 $M_\odot$ BH, its Hawking temperature is below $10^{-7}$ K, far below the current temperature of cosmic microwave background radiation (CMBR). Therefore, the Hawking radiation of BHs will not be effective before most CMB photons are absorbed into BHs or the Universe has expanded to decrease the CMB temperature below that of the Hawking radiation of the BHs. Eventually (after almost an eternal time), the Universe will be in equilibrium between the photons trapped by the BHs and the Hawking radiation at a temperature below $10^{-7}$ K. Such a universe is not much better than a dead universe made of essentially only BHs.

Mathematically, wormholes and white holes may be able to dig out the energy and matter lost in BHs. However, with our current knowledge of physics and astrophysics we do not yet know how wormholes and white holes can be produced in the physical Universe. Although I cannot reject this possibility, this is not favored by me, because I do not want to rescue the Universe from eternal death by relying on unknown physics and astrophysics.

As I discussed briefly in the last section, if Penrose's cosmic censorship is broken, "naked" compact objects may quite possibly exist in the physical Universe (similarly, astrophysical BHs can also be turned into "naked" compact objects), although they have not been identified so far. As shown in Figure 10.7, for "naked" compact objects with extremely small radii, radiative efficiency exceeding 100% is possible. For an external observer, this is equivalent to extracting energy from the "naked" compact object, because globally and on the average energy conservation is required. This situation is similar to the Hawking radiation: the vacuum fluctuations around a BH lead to the escape of particles from just outside the event horizon of a BH, but globally this is equivalent to consuming the energy (mass) of the BH as a result of global energy conservation. Likewise, the energy extracted from the "naked" compact object can be turned into matter through various known physical processes. This scenario is just the re-cycling of the previously accreted matter in BHs. Therefore, with "naked" compact objects, if they do exist, the Universe can indeed be rescued from an eternal death caused by all matter being sucked into BHs. I call this the *"naked" compact object re-cycle conjecture*.

Therefore, my final answer to this question is mixed: *Almost all matter indeed will fall into astrophysical BHs; however, "naked" compact objects can re-cycle matter out, if astrophysical BHs can somehow be turned into "naked" compact objects.*

## SUMMARY, CONCLUDING REMARKS, AND FUTURE OUTLOOKS

In this chapter, I have focused on asking and answering the following questions:

- What is a BH? Answer: There are three types of BHs, namely, *mathematical BHs, physical BHs,* and *astrophysical BHs*. An astrophysical BH, with mass distributed within its event horizon but not concentrated at the singularity point, is not a mathematical BH.
- Can astrophysical BHs be formed in the physical Universe? Answer: Yes, at least this can be done with gravitational collapse.
- How can we prove that what we call astrophysical BHs are really BHs? Answer: Finding direct evidence of the event horizon is not the way to go. Instead, I proposed five criteria that meet the highest standard for recognizing new discoveries in experimental physics and observational astronomy.

- Do we have sufficient evidence to claim the existence of astrophysical BHs in the physical Universe? Answer: Yes, astrophysical BHs have been found at least in some galactic binary systems, at the center of almost every galaxy, and as the central engines of at least some long GRBs.
- Will all matter in the Universe eventually fall into BHs? Answer: Probably "no," because "naked" compact objects, if they do exist with radii smaller than the radii of event horizons for their mass but are not enclosed by event horizons, can rescue the Universe from an eternal death by re-cycling out the matter previously accreted into astrophysical BHs. I call this the *"naked" compact object re-cycle conjecture.*

The main conclusion of this chapter is thus that we have confidence to claim discoveries of astrophysical BHs in the physical Universe with the developments of theoretical calculations and modeling of astrophysical BH formation, accretion, and outflows and the applications of these theories to the ever-increasing amount of astronomical observations of many different types of objects and phenomena. This should be considered as a major verification of Einstein's general relativity, given that the Schwarzschild BH is the very first analytic solution of Einstein's field equations. With this, general relativity has prevailed at the gravity (or curvature) level from the Solar System, where the general relativity correction over the Newtonian gravity is small but still non-negligible, to the vicinity of a BH, where the general relativity effects dominate.

It is then interesting to ask this question: Do we need a quantum theory of gravity in order to further understand astrophysical BHs? My answer is: *Probably no.* There are three reasons for giving this perhaps surprising (and perhaps not welcome) answer:

1. Quantum effects outside astrophysical BHs are unlikely to be important because of their macro scales.
2. No information from matter fallen into an astrophysical BH can be obtained by an external observer.
3. For an external observer, matter inside an astrophysical BH is distributed, but not concentrated at its very center, and thus no physical singularity exists even inside it.

However, a quantum theory of gravity is probably needed to understand the behavior of stellar-mass "naked" compact objects, if Penrose's cosmic censorship is broken, because their densities can be extremely high such that quantum effects will be very important. Therefore, a quantum theory of gravity is needed to understand the "naked" compact object re-cycle conjecture I proposed here.

Finally, I ask one more question: What additional astronomical observations and telescopes are needed to make further progress on our understanding of astrophysical BHs and perhaps also "naked" compact objects? The answer to this question can be extremely long, but I try to be very brief here. Personally, I would like to see two types of major observational breakthroughs:

1. X-ray timing and spectroscopic observations of astrophysical BHs with throughputs at least an order of magnitude higher than the existing *Chandra* and *X-ray Multi-Mirror Mission* (*XMM*)-*Newton* x-ray observatories. This would allow detailed examinations of the structure around astrophysical BHs; detailed mapping; and an understanding of the rich physics of accretion, radiation, and outflows under the extreme physical conditions there, as well as exact measurements of BH masses and spin parameters in many systems. For stellar-mass BHs in binaries, these measurements will help us understand their formation mechanism and evolution of massive stars. For actively accreting supermassive BHs in AGNs, these measurements will be very important for understanding the active interactions between astrophysical BHs and their surrounding environments, as well as the formation, evolution, and growth of their host galaxies. This is a major goal of the *International X-ray Observatory* (*IXO*) (http://ixo.gsfc.nasa.gov/) being proposed in the US, Europe, and Japan; this is also the main scientific objective of the proposed *X-ray Timing and Polarization* (*XTP*) space mission within the Diagnostics of Astro-Oscillation (DAO) Program on China's Space Science Road Map (Guo and Wu, 2009).

2. Imaging astrophysical BHs with telescopes of extremely high angular resolving power. Seeing a hole or a shadow of the size of the event horizon of a BH in any accreting BH system would remove any doubt of the existence of the BH for even the most conservative people. Practically, perhaps the supermassive BH at the center of the Milky Way is the first accreting astrophysical BH to be imaged at an angular resolution capable of resolving its event horizon scale. Sub-millimeter interferometers with very long baselines on the Earth or even in space may be able to do just this in the next decade or so. Theoretically, the best and also technically feasible angular resolution can be achieved with space x-ray interferometer telescope arrays, which can obtain direct images of the smallest x-ray-emitting region just outside the event horizon of a BH, the goal of NASA's proposed BH imager mission *MicroArcsecond X-ray Imaging Mission* (*MAXIM*) (http://maxim.gsfc.nasa.gov/). Imaging astrophysical BHs is also a goal of the Portraits of Astro-Objects (PAO) Program on China's Space Science Road Map (Guo and Wu, 2009).

These two types of observational breakthroughs, to be made with future extremely powerful telescopes in space and on the ground, would revolutionize our understanding of astrophysical BHs. With astrophysical BHs as probes of stellar, galactic, and cosmic evolution, observational and theoretical studies of astrophysical BHs in the physical Universe will play increasingly important roles in astronomy, astrophysics, and fundamental physics.

## ACKNOWLEDGMENTS

I am indebted to Don York for his push, patience, and encouragement on writing this chapter; his many insightful comments and suggestions on the manuscript have clarified several points and improved readability. My student Yuan Liu made a substantial contribution to some of the research work used here (mainly on the question, Can astrophysical BHs be formed in the physical Universe?). I appreciate the discussions (mainly on the question, Will all matter in the Universe eventually fall into BHs?) made with my former student Sumin Tang. My colleague Bifang Liu provided me a literature reference and also made some interesting comments on the radiative efficiency of the advection-dominated accretion flow model. Some of our research results included in this chapter are partially supported by the National Natural Science Foundation of China under Grant Nos. 10821061, 10733010, and 0725313 and by 973 Program of China under Grant No. 2009CB824800.